\documentclass{elsart1p}

\usepackage{graphicx}
\usepackage{amssymb}

\begin{document}

\begin{frontmatter}
    \title{Screened perturbation theory at four loops}

    \author{Lars Kyllingstad}
    \address{
        Department of Physics,
        Norwegian University of Science and Technology,
        N-7491 Trondheim,
        Norway
    }

    \begin{abstract}
        We study the thermodynamics of massless $\phi^4$-theory using
        screened perturbation theory, which is a way to systematically
        reorganise the perturbative series. The free energy and pressure are
        calculated through four loops in a double expansion in powers of
        $g^2$ and $m/T$, where $m$ is a thermal mass of order $gT$.
        The result is truncated at order $g^7$.
        We find that the convergence properties are significantly
        improved compared to the weak-coupling expansion.
    \end{abstract}

\end{frontmatter}

\section{Introduction}
    Recently, Gynther {\it et al}.\ calculated the pressure of massless
    $\phi^4$-theory to order $g^6$ in the  weak-coupling expansion
    \cite{gynther-g6-07}. The weak-coupling pressure for various orders
    of $g$ is shown in Fig.\ \ref{f:pressure}b. Note that it does
    not seem to converge as higher and higher orders are included.
    This is a well-known problem, not only in scalar field theory, but
    also in gauge theories.

    Many methods have been devised to improve upon the convergence of
    this expansion. Among them is \emph{screened perturbation theory} (SPT),
    which was first introduced in thermal field
    theory by Karsch, Patk\'os and Petreczky \cite{karsch}. SPT constitutes a
    reorganisation of the perturbative series so that one resums selected 
    diagrams from all orders of perturbation theory. In the following, we
    will indeed see that using SPT improves the convergence significantly.

    This talk is a brief overview of the calculations and results of
    Ref.\ \cite{andersen-spt-08}.

\section{Screened perturbation theory}
    The Lagrangian density for a massless $\phi^4$-theory is
    \begin{equation}
        \mathcal L =
            \frac{1}{2} (\partial_\mu \phi)(\partial^\mu \phi)
            - \frac{g^2}{24} \phi^4,
    \end{equation}
    where $g$ is the coupling constant. The SPT Lagrangian of this theory
    is defined as
    \begin{equation}
        \mathcal L_{\mathrm{SPT}} =
            \mathcal L_{\mathrm{free}}
            + \mathcal L_{\mathrm{int}},
    \end{equation}
    where
    \begin{equation}
        \mathcal L_{\mathrm{free}} =
            \frac{1}{2} (\partial_\mu \phi)(\partial^\mu \phi)
            - \frac{1}{2} m^2 \phi^2
        \qquad \mathrm{and} \qquad
        \mathcal L_{\mathrm{int}} =
            \frac{1}{2} m_1^2 \phi^2
            - \frac{g^2}{24} \phi^4.
    \end{equation}
    If we set $m_1^2 = m^2$, it is clear that
    $\mathcal L_{\mathrm{SPT}} = \mathcal L$.
    We now take $m_1^2$ to be of order $g^2$ and expand systematically in
    powers of $g^2$. This defines a reorganisation of the perturbative
    series, in which the expansion is about an ideal gas of \emph{massive}
    particles. The mass $m$ has the simple interpretation of a thermal mass.
    A prescription for $m$ is discussed later; for now we take it formally
    to be of order $gT$.

\section{Free energy}
    We calculate the free energy in a double power expansion. That is, first we
    do a loop expansion in powers of $g^2$, and thereafter we expand each
    diagram in powers of $m/T$.

    The inclusion of the mass term in the interaction yields an additional
    Feynman rule not present in the original theory, namely
    $\includegraphics[width=8mm]{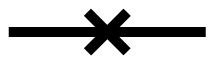} = \frac{1}{2} m_1^2$.
    This is called a \emph{mass insertion}. The free energy can then be written
    as a series of vacuum diagrams,
    \begin{equation}
        \mathcal F =
            \raisebox{-3mm}{\includegraphics[width=8mm]{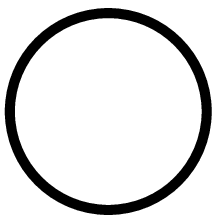}}
            + \raisebox{-3mm}{\includegraphics[width=16mm]{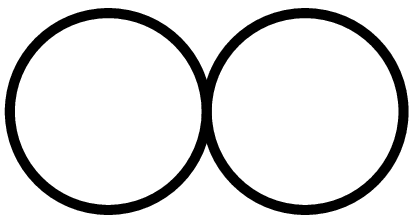}}
            + \raisebox{-3mm}{\includegraphics[width=8mm]{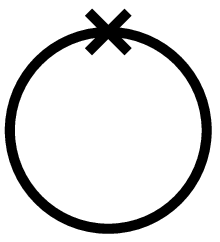}}
            + \raisebox{-3mm}{\includegraphics[width=24mm]{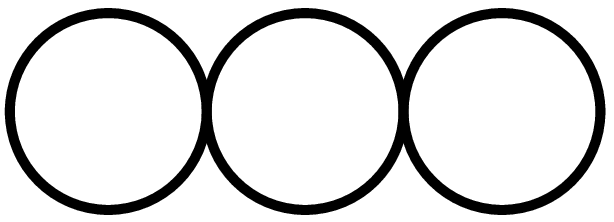}}
            + \raisebox{-3mm}{\includegraphics[width=8mm]{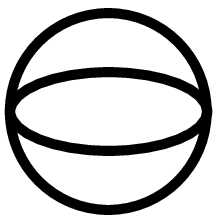}}
            + \raisebox{-3mm}{\includegraphics[width=16mm]{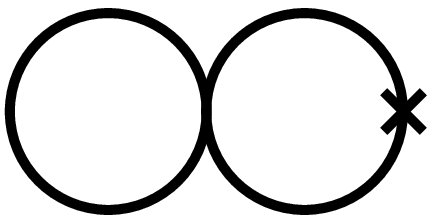}}
            + \cdots,
    \end{equation}
    which we truncate at four loops. Note that mass insertions count as
    loops, since they are of order $g^2$.

    As an example of an $m/T$ expansion, take the one-loop
    diagram with a single mass insertion:
    \begin{equation}
        \mathcal F_{1 \mathrm b}
        \equiv
            \raisebox{-3mm}{\includegraphics[width=8mm]{loop1b}}
        =
            -\frac{1}{2} m_1^2 T \sum_{p_0 = 2 \pi n T}
            \int_p \frac{1}{P^2 + m^2},
    \end{equation}
    There are two momentum scales in this sum-integral; the hard scale,
    which is of order $T$, and the soft scale, of order $gT$.
    The former arises from the nonzero Matsubara frequencies, whereas the
    latter comes from the thermal mass $m$.
    We isolate the contribution from the zeroth Matsubara mode, as it
    only contains the soft scale. This yields
    \begin{equation}
        \mathcal F_{1 \mathrm b} =
            -\frac{1}{2} m_1^2 T
            \left[
                \int_p \frac{1}{p^2 + m^2}
                + \sum_{p_0 \neq 0} \int_p \frac{1}{P^2 + m^2}
            \right].
    \end{equation}
    Since $m \ll P$ in the second term, we can expand it in a geometric series:
    \begin{equation}
        \mathcal F_{1 \mathrm b} =
            -\frac{1}{2} m_1^2 T
            \left[
                \int_p \frac{1}{p^2 + m^2}
                + \sum_{p_0 \neq 0} \int_p \frac{1}{P^2}
                \left(
                    \frac{m^2}{P^2}
                    + \frac{m^4}{P^4}
                    + \cdots
                \right)
            \right].
    \end{equation}
    The mass can now be taken outside the sum-integral in each term, and the
    result is a series of easily-evaluable massless sum-integrals.
    Finally, the results are truncated at $g^7$.

\section{The tadpole mass}
    The pressure of the original theory is obtained in the limit where the
    two masses are equal, and is defined as
    \begin{equation}
        \mathcal P = -\mathcal F |_{m_1^2 = m^2}.
    \end{equation}
    The parameter $m$ in screened perturbation theory is completely arbitrary,
    and if we were able to include \emph{all} loop orders, the result
    would indeed be independent of $m$.
    To complete the calculation we must instead find a prescription for
    $m$ which is physically meaningful.

    The simplest choice is the tadpole,
    \begin{equation}
        m^2
        =
            \raisebox{-3mm}{\includegraphics[width=16mm]{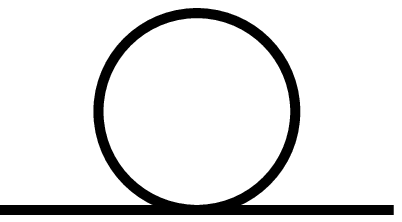}}
        =
            \frac{1}{2} g^2 T \sum_{p_0}
            \int_p \frac{1}{P^2 + m^2}.
        \label{eq:tadpole}
    \end{equation}
    In the weak-coupling limit the propagator in the loop is massless,
    and Eq.\ (\ref{eq:tadpole}) reduces to
    \begin{equation}
        m^2 = \frac{g^2 T^2}{24}.
    \end{equation}
    Using this value for the mass one obtains the weak-coupling pressure,
    shown in Fig.\ \ref{f:pressure}b. Our result through order $g^6$
    agrees with the $N=1$ result in Ref.\ \cite{gynther-g6-07}.

    We can generalise this to higher loop orders by taking $m$ to be the
    \emph{tadpole mass},
    \begin{equation}
        m^2 =
            \includegraphics[width=15mm]{tadpole0a}
            + \includegraphics[width=15mm]{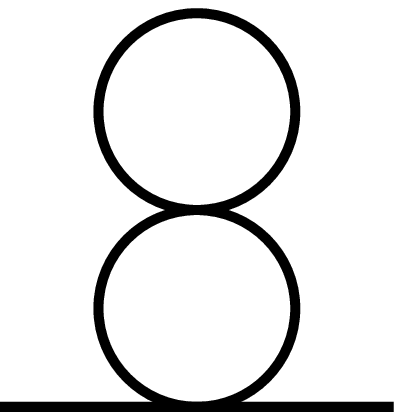}
            + \includegraphics[width=15mm]{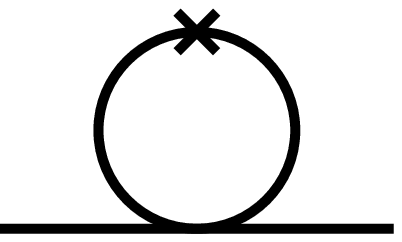}
            + \cdots
        =
            g^2 \left.
            \frac{\partial \mathcal F}{\partial (m^2)}
            \right|_{m_1^2 = m^2}.
        \label{eq:tadpole_mass}
    \end{equation}
    With this choice, $m$ is well-defined at all loop orders. Since the
    propagators in Eq.\ (\ref{eq:tadpole_mass}) are massive as well, it
    means that in calculating the pressure we are doing a selective resummation
    of diagrams from all orders of perturbation theory.

\section{Results}
    \begin{figure}
        \center
        \includegraphics[width=12cm]{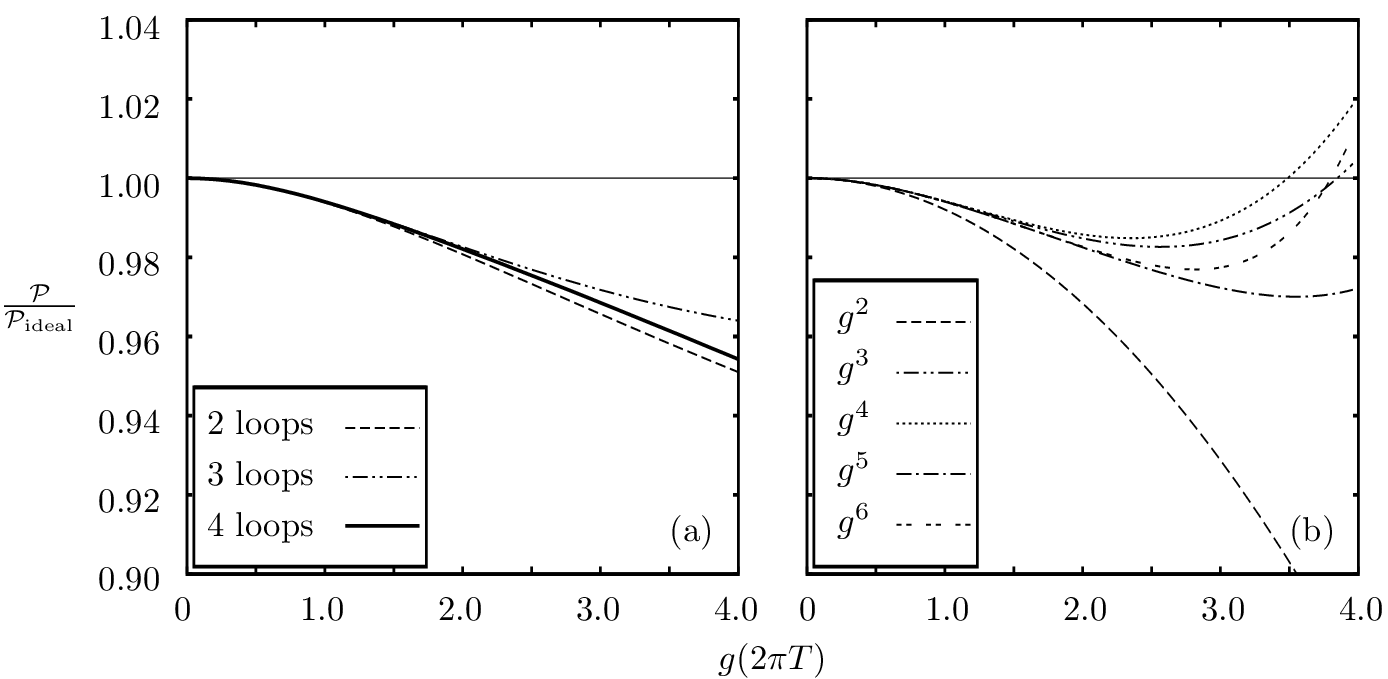}
        \caption{
            (a) Pressure normalised to $\mathcal P_{\mathrm{ideal}}$
                through $g^7$ for various loop orders in SPT.
            (b) Weak-coupling pressure through various orders of $g$
                \cite{gynther-g6-07,arnold-threeloop-94,parwani-g5-95,braaten-g5-95}.
        }
        \label{f:pressure}
    \end{figure}

    Fig.\ \ref{f:pressure}a shows the SPT pressure truncated at various loop
    orders. The two- and three-loop results are indistinguishable from the exact
    numerical results found in Ref.\ \cite{andersen-spt-01}. Convergence
    is rapid---in the two-loop case terms of order $g^5$--$g^7$ are negligible,
    while at three loops one can neglect terms of order $g^7$.

    There are no exact numerical data available for comparison with our
    result at four loops, but experience with lower loop orders indicates that
    this is indeed a good approximation. This can, however, only be confirmed by
    calculating the pressure through $g^8$.

\section{Summary and outlook}
    We have calculated the pressure of a massless $\phi^4$ theory using
    screened perturbation theory. As Fig.\ \ref{f:pressure} shows,
    the successive approximations in SPT seem a lot more stable
    than in the weak-coupling expansion. The apparent improved convergence seems
    to be linked to the fact that SPT is basically an expansion about an
    ideal gas of massive particles, instead of an expansion about an ideal
    gas of massless particles, which is the case for the weak-coupling
    expansion.
    
    Note that in Fig.\ \ref{f:pressure}b, only terms through order $g^6$ in
    the weak-coupling expansion have been included.
    This is because part of the $g^7$-contribution arises from
    five-loop vacuum diagrams which aren't considered in
    Ref.\ \cite{andersen-spt-08}.
    Evaluation of the free energy to order $g^7$ is work currently in
    progress \cite{andersen-g7-0x}.

\vspace{5mm}
\noindent {\bf Acknowledgements}
\vspace{4mm}

    This work was done in collaboration with Jens O. Andersen.
    The author would like to thank the organising commitee of
    SEWM08 for an interesting and stimulating conference.

\end{document}